\documentclass{article}

\usepackage{microtype}
\usepackage{graphicx}
\usepackage{subfigure}
\usepackage{booktabs} 
\usepackage{amsmath}
\usepackage{float}
\usepackage{xcolor}
\usepackage{amsfonts}

\usepackage{todonotes}

\usepackage{hyperref}

\usepackage[accepted]{icml2019}

\icmltitlerunning{Extending Deep Learning Models for Limit Order Books to Quantile Regression}

\begin{document}

\twocolumn[
\icmltitle{Extending Deep Learning Models for Limit Order Books to Quantile Regression}



\icmlsetsymbol{equal}{*}

\begin{icmlauthorlist}
\icmlauthor{Zihao Zhang}{to}
\icmlauthor{Stefan Zohren}{to}
\icmlauthor{Stephen Roberts}{to}
\end{icmlauthorlist}

\icmlaffiliation{to}{Oxford-Man Institute of Quantitative Finance, Department of Engineering Science, University of Oxford, Oxford, United Kingdom}

\icmlcorrespondingauthor{Zihao Zhang}{zihao@robots.ox.ac.uk}

\icmlkeywords{Deep Learning, Quantile Regression, Limit Order Book, Time-Series Analysis,  ICML}

\vskip 0.3in
]


\printAffiliationsAndNotice{}  

\begin{abstract}
    We showcase how Quantile Regression (QR) can be applied to forecast financial returns using Limit Order Books (LOBs), the canonical data source of high-frequency financial time-series. We develop a deep learning architecture that simultaneously models the return quantiles for both buy and sell positions. We test our model over millions of LOB updates across multiple different instruments on the London Stock Exchange. Our results suggest that the proposed network not only delivers excellent performance but also provides improved prediction robustness by combining quantile estimates. 
\end{abstract}

\section{Introduction} 
\label{introduction}

Traditional time-series modelling is often dominated by Markov-like models with stochastic driving terms such as the vector autoregressive model (VAR) \cite{zivot2006vector}. These models make strong parametric assumptions to the functional form of the predictive model (in particular the AR family) and also require the target time-series to be stationary. However, financial price (and volume) rarely conform to these assumptions and even returns, the first order differences of prices, are rarely stationary \cite{cont1999statistical}. Deep learning has gained popularity in financial modelling since they are not constrained by the above assumptions (see \cite{tsantekidis2017forecasting, tsantekidis2017using} for some examples). Modern deep learning architectures also allow one to tailor the loss function, as is demonstrated in \cite{lim2019enhancing} where the Sharpe ratio is directly  maximised as a cost function. 

Deep networks often require a large number of observations to calibrate weights and this property fits nicely with financial applications that utilise high frequency microstructure data. Nowadays, billions of market data are generated everyday and most of them are recorded in Limit Order Books (LOBs) \cite{parlour2008limit, bouchaud2018trades}.
A LOB is a record of all unmatched orders of a given instrument in a market comprising of levels at different prices containing resting limit orders to sell and buy, also called ask and bid orders. A bid (ask) order is an order to buy (sell) an asset at or below (above) a specified price. We can consider LOBs as the most granular financial data as a LOB represents the demand and supply of a given instrument at any moment in time. 


In our previous works \cite{zhang2018bdlob, zhang2019deeplob}, we demonstrate that deep learning models can deliver improved predictive performance, in comparison with standard methods when modelling LOB data. One of the important contributions of deep learning is the ability to automate the process of feature extraction. In the work of \cite{sirignano2018universal} as well as our own, it is demonstrated that such models can extract representative features that are related to the demand and supply in the order book. Given billions of market quotes from LOBs, this approach has proved to be more effective than algorithms that rely on hand-crafted features. Features derived from a human-centric understanding of a process (such as ``moving-average crossover'') do not guarantee best performance with respect to the target function, as is also seen in \cite{lim2019enhancing}. Further, in complex non-stationary environments such as finance, it is far from trivial to select informative features by hand, even after years of work in the industry.

In this work, we utilise deep neural networks and Quantile Regression (QR) \cite{koenker2001quantile} to model the returns from LOBs.  QR is particularly useful for financial time-series as it models the conditional quantiles of a response. Returns are in general heterogeneous, highly peaked and have fat tails compared to a normal distribution \cite{cont1999statistical}. A point estimation is really not enough to describe the full distribution of returns and we can obtain considerably more information by estimating multiple quantiles. We regard Quantile Regression as able to provide valuable, non-stationary, extra information regarding risk exposure.

\begin{figure}[t]
\vskip 0.2in
\begin{center}
\centerline{\includegraphics[width=3.3in, height=3.4in]{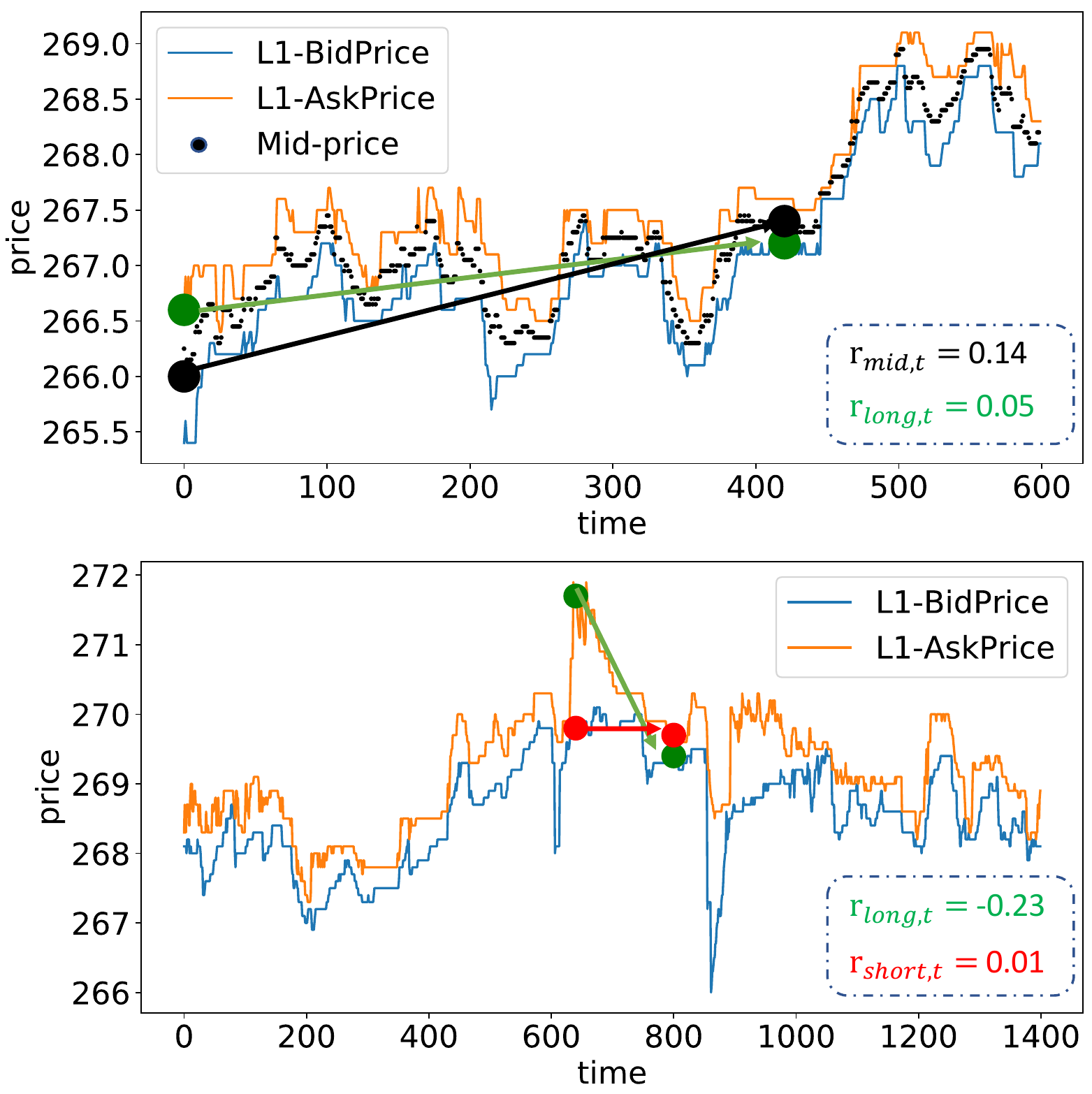}}
\caption{\textbf{Top:} returns $r$ obtained from using mid-price or first level prices from LOBs. \textbf{Bottom:} returns obtained from long and short positions at a given time stamp.}
\label{fig:spread}
\end{center}
\vskip -0.2in
\end{figure}

Unlike most literature, where mid-price is used to represent financial time-series, we define returns by directly using first level prices from LOBs. Modelling mid-price is appropriate if we are using daily data, but it is improper for intraday strategies as we ignore spreads which are the differences between best ask and bid price.
The upper plot of Figure~\ref{fig:spread} illustrates the effects of spreads as transaction costs on returns. 
The return, $r_{mid,t}$, from using mid-price is about three times higher than the actual return, $r_{long,t}$, that we can obtain (we have to buy from ask sides and sell from bid sides for aggressively entering or exiting positions). Further, intraday strategies often involve both long and short positions to increase profitability. However, returns from these two positions are not symmetric at a given time stamp. We observe this at the bottom of Figure~\ref{fig:spread}. A return, $r_{long,t}$, is -0.23 from a long position, but it does not imply a profit of 0.23 by taking a short position ($r_{short,t} = 0.01$). Indeed, the two return series are statistically different under Kolmogorov-Smirnov and Wilcoxon signed-rank tests. This discrepancy comes from changing spreads and indicates that separate models are required to estimate returns for different positions.

\textit{Our contributions:} We propose a network architecture that can simultaneously estimate multiple return quantiles from both buy and sell positions by training with different Quantile Loss functions. Our model consists of a block of convolutional layers and multiple LSTM branches to estimate different quantiles. The convolutional block, as a feature extraction mechanism, processes raw limit order book data and LSTM layers are used to capture time dependencies among the resulting feature maps. We show that this method delivers better predictive performance than other popular machine learning algorithms. Also, better performance can be achieved by combining estimates from different quantiles. 

\section{Data Description and Returns}
\label{data}

Our dataset consists of one year full-resolution LOB data for five of the most liquid stocks listed on the London Stock Exchange (LSE), namely, Lloyds Bank, Barclays, Tesco, BT and Vodafone. The data spans all trading days from the 3rd of January 2017 to the 24th of December 2017 and only normal trading periods (between 08:30:00 and 16:30:00) are included. We take price and volume for 10 levels on both ask and bid sides of a LOB so there are 40 features at each timestamp. Overall,  our dataset has more than 134 million observations and there are, on average, 150,000 events per day per stock. The first 6 months are used as training data, the next 3 months as validation data and the last 3 months as test data. In the context of high-frequency data, 3 months test data corresponds to millions of observations and therefore provides sufficient scope for testing model performance and robustness. 

\begin{figure}[htb]
\vskip 0.2in
\begin{center}
\centerline{\includegraphics[width=3.3in, height=1.6in]{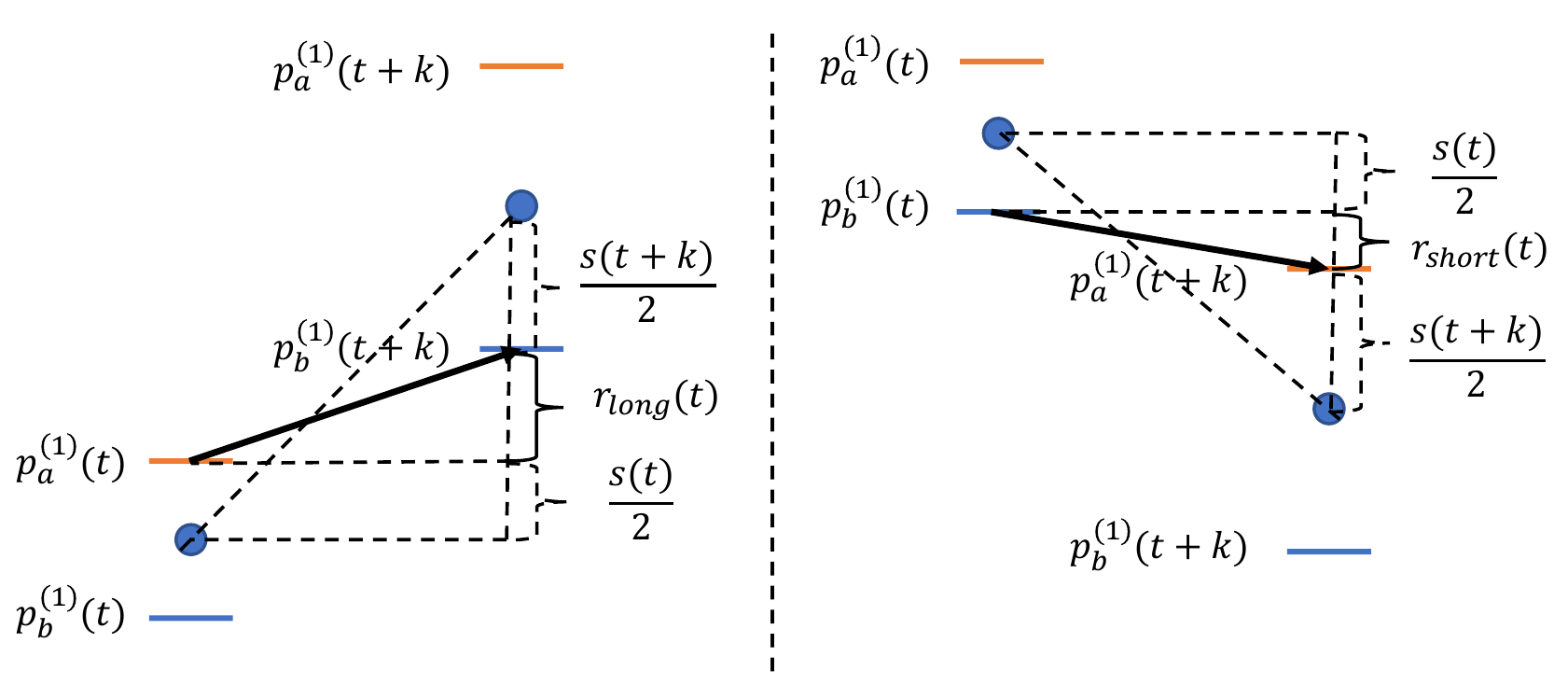}}
\caption{A schematic description of Equation~\eqref{eq:return1}. $p_a^{(1)}(t)$ and $p_b^{(1)}(t)$ represent best ask and bid prices at time $t$. \textbf{Left:} Shows a return from a long position; \textbf{Right:} Shows a return from a short position.}
\label{fig:return_decompose}
\end{center}
\vskip -0.2in
\end{figure}

We prepare our input data using the procedure outlined in our previous work \cite{zhang2019deeplob}. We define returns by taking spreads into account. A return, $r(t)$, at time $t$, both long or short, can be decomposed as:
\begin{equation} \label{eq:return1}
\!\!r_i(t)\! =\!\frac{\Delta p_i(t)}{p_{mid}(t)},    \Delta p_i(t) \!=\! z_i(t)\Delta m(t)\! -\!\frac{s(t)\!+\!s(t\!+\!k)}{2} 
\end{equation}
where 
\begin{equation}
    \begin{split}
    z_i(t) &= \left \{
    							\begin{tabular}{cc}
    							  1,  & if $i=\mathrm{long}$ \\
							  -1, & if $i=\mathrm{short}$
							 \end{tabular}  \right. \\
    \Delta m(t) &= p_{mid}(t+k) - p_{mid}(t) \\
    p_{mid}(t) &= \frac{p_{ask}^{(1)}(t) + p_{bid}^{(1)}(t)}{2}
    \end{split}
\end{equation}
and $k$ is the predicted horizon, $s(t)$ and $s(t+k)$ are spreads at time $t$ and $t+k$. We denote the first level ask and bid price as $p_{ask}^{(1)}(t)$ and $p_{bid}^{(1)}(t)$.  A schematic description of Equation~\eqref{eq:return1} is given in Figure~\ref{fig:return_decompose} -- we cross half the spread now, follow the mid price and cross the other half later. As we already observe the current spread $s(t)$, there is no need to model it and we take it out from \eqref{eq:return1}, so the return $r_i'(t)$ of interest is defined as:
\begin{equation} \label{eq:return2}
r_i'(t)= \frac{\Delta p_i'(t) }{p_{mid}(t)}, \quad \Delta p_i'(t) = \Delta p_i(t) + s(t).
\end{equation}


Note that Equation~\eqref{eq:return2} can be also written as:
\begin{equation} \label{eq:return3}
    r_i'(t) =  z_i(t) r_{mid}(t) -  r_{spread}(t)/2
\end{equation}
where $r_{mid} = \Delta m(t)/p_{mid}(t)$ and $r_{spread}(t) = (s(t+k)-s(t))/p_{mid}(t)$. Instead of modelling ($r'_{long}(t), r'_{short}(t)$), we can also estimate quantiles for mid-price change and spread change ($ r_{mid}(t) , r_{spread}(t) $). 

\section{Methods and Network Architecture}
\label{methods}

\subsection{Quantile Regression}

In Quantile Regression (QR), we predict the conditional quantile of the target distribution, i.e. $\mathbb{P}(r\leq r^{(\tau)} | x) = \tau$ , where $\tau$ is the quantile of interest and $x$ is the input. We can obtain the estimates by minimizing the Quantile Loss (QL) with respect to $r^{(\tau)}$, leading to the $\tau$th quantile:
\begin{eqnarray}
    L_{\tau}(r, \hat{r}^{\tau}) &=& \sum_{t: r(t) < \hat{r}^{\tau}(t)} (\tau - 1)|r(t) - \hat{r}^{\tau}(t)| +  \nonumber\\
    &&\sum_{t: r(t) > \hat{r}^{\tau}(t)} \tau |r(t) - \hat{r}^{\tau}(t)|
\end{eqnarray}
where $r(t)$ is the observation at time $t$ and $\hat{r}^{\tau}(t)$ is the predicted value. The common used mean absolute error is equivalent to QL with $\tau = 0.5$. Note that each quantile has its own QL function and, in order to obtain multiple quantiles, we need separate models to estimate each of them. Our network solves this constraint by training with multiple QLs to model all quantiles of interest simultaneously. 

QR has a common problem known as quantile crossing as quantile curves can cross each other, leading to an invalid distribution for the response, e.g. the predicted 90th percentile of the response is smaller than the 80th percentile which is impossible. We follow the work of \cite{chernozhukov2010quantile} to rearrange the original estimated non-monotone curve into a monotone rearranged curve.

\subsection{Network Architecture}
We propose a neural network architecture that simultaneously models different quantiles of returns from both long and short positions. This model is an extension of our previous work \cite{zhang2019deeplob} and, for a detailed discussion on how initial layers are constructed, please find more information there. We denote this new architecture as DeepLOB-QR. 

As each quantile has its own loss function, if we are interested in 3 quantiles for returns from each position, we would need to estimate 6 separate models. This is computationally demanding for LOBs data as we have millions of them in a single day. Further, each of the models is essentially estimating the ``same'' underlying quantity (long or short return), just different quantiles of it. We would thus expect that there are common features that can contribute to the estimation of all models and it is wasteful to estimate them separately.

DeepLOB-QR is designed to solve the above constraints by using a common convolutional block and branching out several LSTM layers to model different interested quantiles. The ``main input'' takes raw LOBs data to extract features that can modulate relationships between demand and supply of an instrument. The two auxiliary inputs take past returns from long and short positions.
We isolate LOBs data and past returns here because one important property of convolutional layer is parameter sharing and price and volume series have different dynamics compared to returns. Overall, this is a multi-input and multi-ouput setup but trained using different loss functions (6 QLs in our case). The last parallel LSTM layers (LSTM$@32$) are only trained using their corresponding losses, while each of the two LSTM$@64$ layers is trained using 3 losses and the convolutional block is trained using all losses. 

\begin{figure}[t]
\vskip 0.2in
\begin{center}
\centerline{\includegraphics[width=3.3in, height=1.7in]{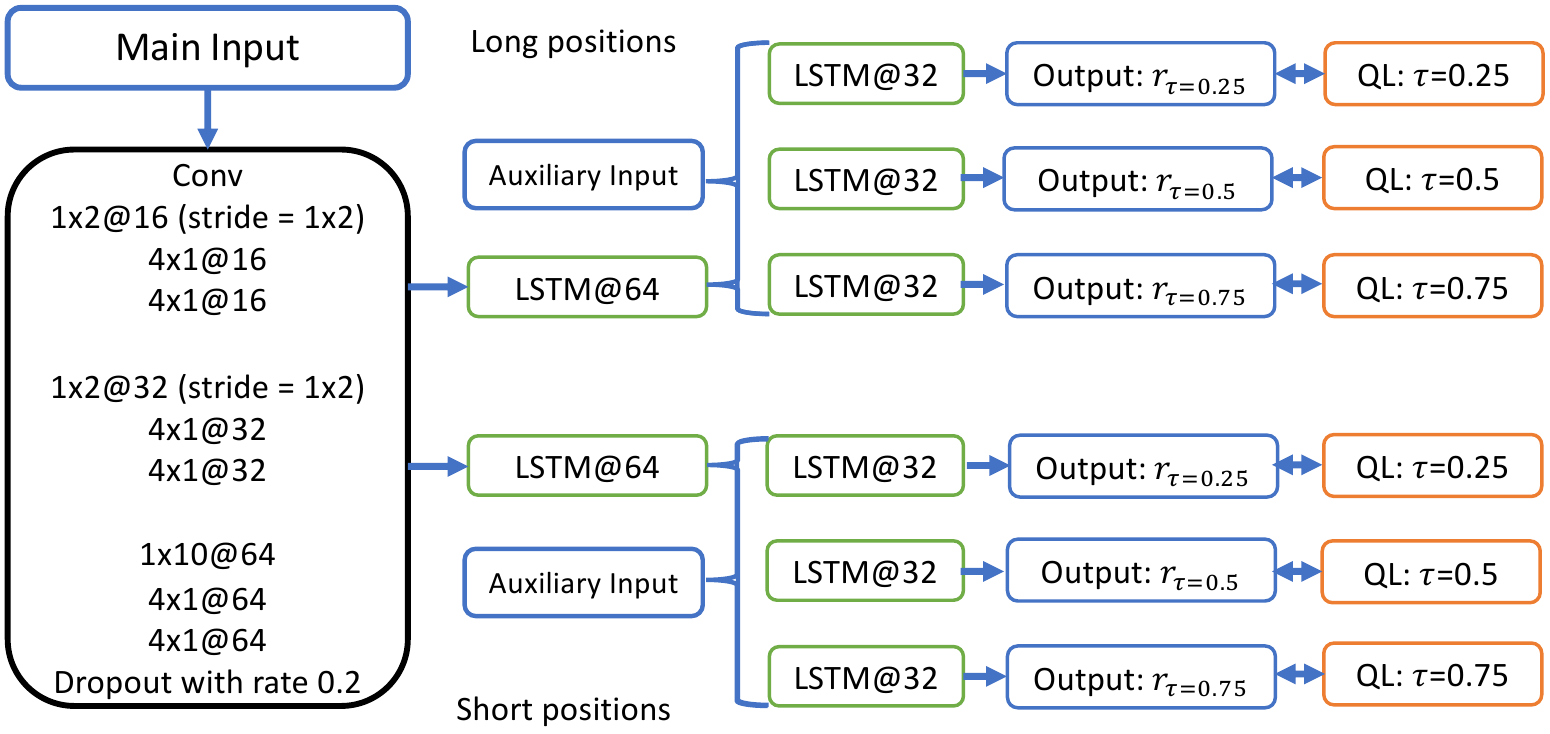}}
\caption{Model architecture for DeepLOB-QR. Here $1 \times 2@16$ represents a convolutional layer with 16 filters of size $(1\times 2)$ . QL: $\tau$ represents the quantile loss at quantile $\tau$.}
\label{fig:model}
\end{center}
\vskip -0.2in
\end{figure}

\subsection{Forecast Combination}

The works of \cite{ rapach2010out, meligkotsidou2014quantile} suggest that combinations of individual quantile estimates can form a much robuster point estimation and help reduce prediction uncertainty. After obtaining quantile estimates $\hat{r}^{(\tau)}(t)$, $\tau \in \mathcal{S}$ where $\mathcal{S}$ denotes the set of considered quantiles, we can combine them as:
\begin{equation}
\hat{r}(t) = \sum_{\tau \in \mathcal{S}} \pi^{(\tau)} \hat{r}^{(\tau)}(t), \quad \sum_{\tau \in \mathcal{S}} \pi^{(\tau)}=1
\end{equation}
where the weights $\pi^{(\tau)}$ represents the probability assigned to the prediction of quantile $\tau$. This estimator, as a linear combination of order statistics, forms a point estimation of the central location of a distribution based on small sets of quantile estimates. We can reflect our beliefs on how each quantile estimate affects the central location by adjusting the corresponding weights. 

The simplest method of estimating the weights is to use a fixed weighting scheme. We can also form a constrained optimzation problem \cite{meligkotsidou2014quantile} to find an optimal combination of quantile estimates:
\begin{equation}
    \mathrm{\mathbf{\pi}} = \arg \underset{\mathrm{\mathbf{\pi}}}{\min} \ E_t [r(t-1) -  \sum_{\tau \in \mathcal{S}} \pi^{(\tau)} \hat{r}^{(\tau)}(t-1) ]^2,
    \label{eq:dfc}
\end{equation}
where $\mathrm{\mathbf{\pi}} = [\pi^{(\tau)}]_{\tau \in \mathcal{S}}$ and $\sum_{\tau \in \mathcal{S}} \pi^{(\tau)}=1$. This procedure hence treats quantile estimation as constrained regression, where the weights are non-negative and sum to unity. 

\section{Experiments and Results}
\label{results}

As in Figure~\ref{fig:model}, we predict three quantiles (0.25, 0.5 and 0.75) for returns from long positions and same three quantiles for short positions, totalling 6 QLs. Our prediction horizon, $k$, is 100 steps into the future. In order to compare with other methods, we assess how different models estimate the central location of a response. We denote estimates of the 0.5 quantile from our model as DeepLOB-QR and estimates obtained from the combination scheme (Equation~\eqref{eq:dfc}) as DeepLOB-QR(C). We compare to four other models: an autoregressive model (AR), a generalised linear model (GLR), a support a vector regression (SVR) and a neural network with multiple fully connected layers (MLP). 

\begin{figure}[htb]
\vskip 0.2in
\begin{center}
\centerline{\includegraphics[width=3.3in, height=2.1in]{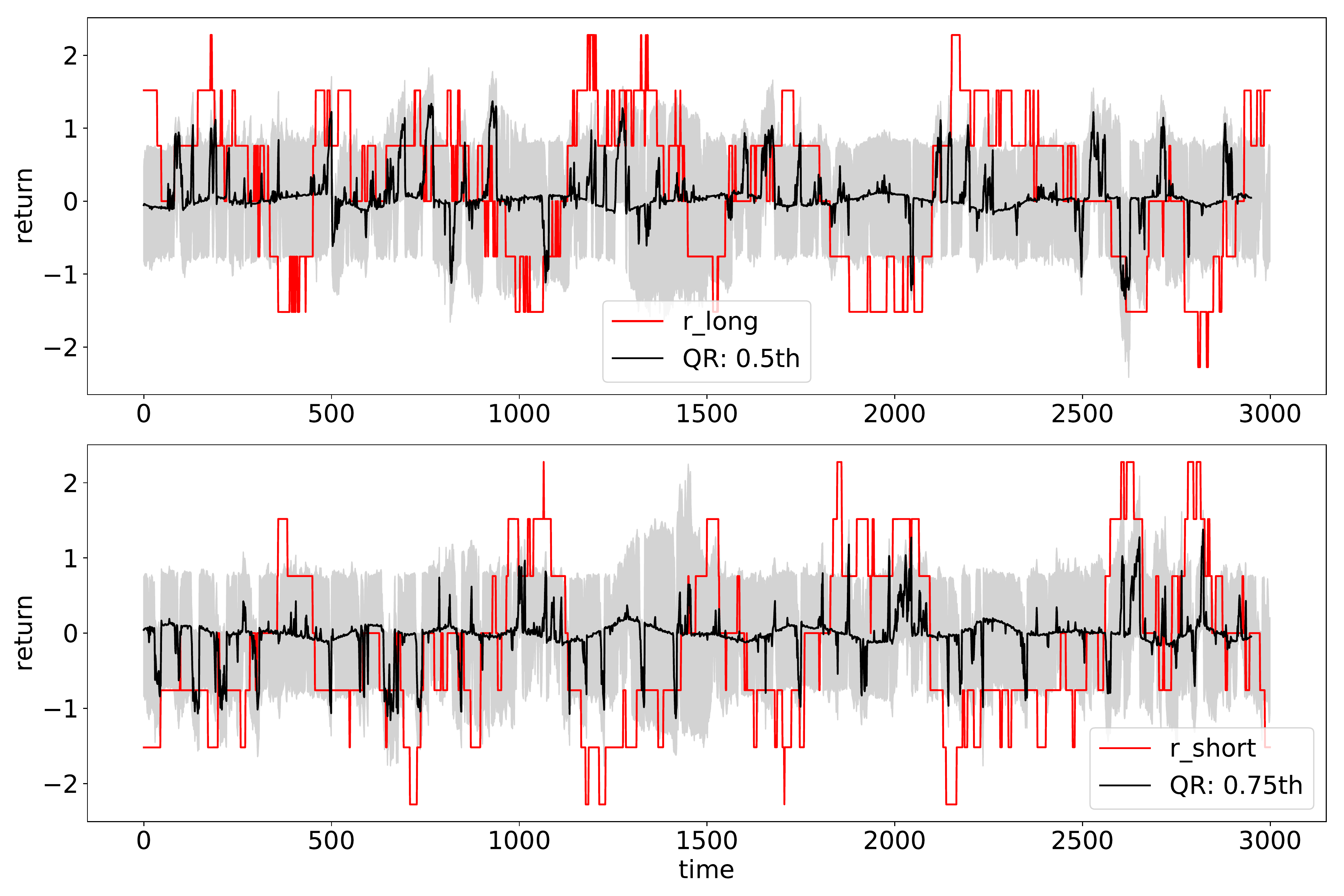}}
\caption{\textbf{Top:} Real returns from long positions is in red and 0.5 quantile estimate is in black. The upper boundary of grey shading represents 0.75 quantile estimate and the lower boundary is for 0.25 quantile estimate; \textbf{Bottom:} Real returns and quantile estimates for short positions.}
\label{fig:quantile_estimate}
\end{center}
\vskip -0.2in
\end{figure}

Table~\ref{result-table} shows the results of our model compared to other methods. Performance is measured by mean absolute error (MAE), mean squared error (MSE), median absolute error (MEAE) and $R^2$ score. All errors, except $R^2$ score, are divided by errors from a repetitive model\footnote{A repetitive model means using current observations as future predictions, a zero-intelligence model.}, therefore, we understand how much better a model is compared to a zero-intelligence method. DeepLOB-QR achieves the smallest errors and the highest $R^2$ for modelling returns from both long and short positions. Also, our results suggest that better prediction results can be obtained from forecast combinations (DeepLOB-QR(C)). However, as the prediction horizon ($k$) increases, the problem becomes more difficult and we obtain worse performance, as shown in Table~\ref{result-horizon}. To visualise how quantile estimates form a prediction interval, we plot them against the true observations in Figure~\ref{fig:quantile_estimate}.

\begin{table}[t]
\caption{Experiment results for the LSE dataset.}
\label{result-table}
\vskip 0.15in
\begin{center}
\begin{small}
\begin{sc}
\begin{tabular}{l|cccc}
\toprule
Model              & MAE      & MSE     & MeAE    & $R^2$    \\
\midrule
\multicolumn{5}{c}{Metrics for returns from long positions}  \\
\midrule
AR                    &0.705     &0.466     &0.793     &0.004       \\
GLR                 &0.719          &0.556         & 0.707        &-0.188          \\
SVR                 &0.802          &0.498         &0.755         &-0.063          \\
MLP                 &0.717          &0.466         &0.708         &0.008          \\
DeepLOB-QR         &0.701          &0.462         &0.710         &0.010          \\
DeepLOB-QR(C)      &\textbf{0.701}          &\textbf{0.461}         &\textbf{0.702}         &\textbf{0.014}          \\
\midrule
\multicolumn{5}{c}{Metrics for returns from short positions} \\
\midrule
AR                    & 0.708        & 0.498        &0.791         &0.004          \\
GLR                 &0.736          &0.528         &0.709         &-0.055          \\
SVR                 &0.738          &0.548         &0.708        &-0.096          \\
MLP                 &0.726          &0.510         &0.708         & 0.002         \\
DeepLOB-QR         &0.705          &0.501         & 0.720        &0.013          \\
DeepLOB-QR(C)      &\textbf{0.702}          &\textbf{0.494}         &\textbf{0.699}         &\textbf{0.015}         \\
\bottomrule
\end{tabular}
\end{sc}
\end{small}
\end{center}
\vskip -0.1in
\end{table}

\begin{table}[t]
\caption{Experiment results for different prediction horizons $k$.}
\label{result-horizon}
\vskip 0.15in
\begin{center}
\begin{small}
\begin{sc}
\begin{tabular}{l|cccc}
\toprule
Model              & MAE      & MSE     & MeAE    & $R^2$    \\
\midrule
\multicolumn{5}{c}{Metrics for Returns from Long Positions}  \\
\midrule
DeepLOB-QR$_{k=50}$        &       0.626 &       0.524&       0.521&    0.025    \\
DeepLOB-QR(C)$_{k=50}$     &      0.634 &      0.514 &       0.513&    0.042    \\
DeepLOB-QR$_{k=100}$       &0.701          &0.462         &0.710         &0.010          \\
DeepLOB-QR(C)$_{k=100}$    &    0.701   &0.461         &0.702         &0.014          \\
DeepLOB-QR$_{k=200}$      &       0.701&       0.417&       0.864&      0.010  \\
DeepLOB-QR(C)$_{k=200}$    &      0.719 &       0.419&     0.841  &   0.011    \\
\midrule
\multicolumn{5}{c}{Metrics for Returns from Short Positions} \\
\midrule
DeepLOB-QR$_{k=50}$        &       0.624&       0.493&      0.494 &      0.024  \\
DeepLOB-QR(C)$_{k=50}$     &     0.644  &      0.483 &     0.501  &       0.044 \\
DeepLOB-QR$_{k=100}$       &0.705          &0.501         & 0.720        &0.013          \\
DeepLOB-QR(C)$_{k=100}$    &0.702          &0.494         &0.699         &0.015        \\
DeepLOB-QR$_{k=200}$      &       0.700&      0.428 &    0.864   &       0.012 \\
DeepLOB-QR(C)$_{k=200}$   &       0.718&    0.430   &     0.837  &  0.011      \\
\bottomrule
\end{tabular}
\end{sc}
\end{small}
\end{center}
\vskip -0.1in
\end{table}

\section{Conclusion}

In the context of time-series from high-frequency limit order book (LOB) data we show that quantile regression (QR) can provide us with prediction uncertainty by forming a confidence interval and better predictve performance can be obtained by combining multiple quantile estimates. A promising direction for future work is to design trading strategies based on derived signals. We can combine QR with Reinforcement Learning and use uncertainty information to tackle the exploration and exploitation problem. 




\end{document}